\documentclass[
	12pt,				% tamanho da fonte
	oneside,			% para impressão em recto e verso. Oposto a oneside
	a4paper,			% tamanho do papel.
	aps,pre,superscriptaddress,notitlepage,
	nofootinbib,
	english,			% idioma adicional para hifenização
	]{revtex4-1}

\usepackage{lmodern}			% Usa a fonte Latin Modern			
\usepackage[T1]{fontenc}		% Selecao de codigos de fonte.
\usepackage[utf8]{inputenc}		% Codificacao do documento (conversão automática dos acentos)
\usepackage{lastpage}			% Usado pela Ficha catalográfica
\usepackage{indentfirst}		% Indenta o primeiro parágrafo de cada seção.
\usepackage{color}			% Controle das cores
\usepackage{graphicx}			% Inclusão de gráficos
\usepackage{microtype} 			% para melhorias de justificação
\usepackage{amsmath,amssymb,amsfonts}	%Extends the Computer Modern family (of fonts) with additional symbols and alphabets for use in math contexts.
\usepackage{dsfont}			% Double-struck letters: for the natural/integers/real/complex numbers, identity.
\usepackage{float}      % Floats are containers for things that cannot be broken over a page. By default "table" and "figure" are floats
\usepackage{pgfplots}
\usepackage[version=4]{mhchem}			%formulas quimicas
\usepackage{hyperref}				%Links em equações e referências.
\usepackage[left=2.5cm, right=2cm, top=2cm]{geometry} %Controle de tamanho de margens.
\usepackage{etoolbox}
\usepackage{xcolor}
\usepackage[export]{adjustbox}

\newcommand{\overbar}[1]{\mkern 3.0mu\overline{\mkern-3.0mu#1\mkern-3.0mu}\mkern 3.0mu}   % barra sobre letras

\definecolor{blue}{RGB}{41,5,195}

\begin{document}

% Retira espaço extra obsoleto entre as frases.
\frenchspacing 

% Informações de título e autores
\title{Quantum Ising model in a period-2 modulated transverse field}
\author{Adalberto D. Varizi}
\email[Corresponding author: ]{adalbertovarizi@gmail.com}
\affiliation{Departamento de F\'isica, Universidade Federal de Minas Gerais, Belo Horizonte, Minas Gerais 31270-901, Brazil}
\author{Raphael C. Drumond}
\affiliation{Departamento de Matem\'atica, Universidade Federal de Minas Gerais, Belo Horizonte, Minas Gerais 31270-901, Brazil}
\date{\today}
% resumo
\begin{abstract}
We study a finite spin-$\frac{1}{2}$ Ising chain with a spatially alternating transverse field of period 2. By means of a Jordan-Wigner transformation for even and odd sites, we are able to map it into a one-dimensional model of free fermions. We determine the ground-state energies in the positive- and negative-parity subspaces (subspaces with an even or odd total number of down spins, respectively) and compare them in order to establish the ground-state energy for the entire Hamiltonian. We derive closed-form expressions for this energy gap between the different parity subspaces and analyze its behavior and dependence on the system size in the various regimes of the applied field. Finally, we suggest an expression for the correlation length of such a model that is consistent with the various values found in the literature for its behavior in the vicinity of critical points.
\end{abstract}
% Imprime título
\maketitle
% ELEMENTOS TEXTUAIS
% ----------------------------------------------------------

\section{Introduction}

The spin-$\frac{1}{2}$ quantum Ising model in a transverse field is an archetypal model for the theory of quantum phase transitions and magnetism \cite{qpt}. It can be realized experimentally in the ferromagnet \ce{CoNb2O6} \cite{Coldea} and also in trapped cold atoms \cite{Simon}. It is commonly used as a test model in numerical techniques \cite{Porras,Orus} and to elucidate or test new concepts, such as decoherence in open quantum systems \cite{Haikka}, the role of entanglement in phase transitions \cite{Brandao}, and in quantum thermodynamics definitions of work \cite{Cosco,Fusco}. A standard description of its diagonalization procedure, ground- and excited-state determination, and derivation of its free energy and other thermodynamic properties may be found in \cite{qpt,Pfeuty,Lieb,Katsura}. A thorough discussion about the diagonalization and the ground-state determination in a \textit{finite} Ising model is given in Ref. \cite{IsingModel}. There the authors show that, after a Jordan-Wigner transformation, the model's Hamiltonian may be broken into and diagonalized in independent subspaces of positive and negative parity - i.e., subspaces with even or odd numbers of quasiparticles, respectively. A closed-form expression for the gap between the positive- and negative-parity ground-states is also given and its asymptotic behavior analyzed.

An interesting modification of the Ising model was considered in Ref. \cite{Derzhko}, where a regular alternation of the type $g-(-1)^j\delta_g$, where $j$ is a spin site, was introduced in the transverse field and its ground-state and low-temperature properties are discussed for an infinite system. They showed that for different values of the transverse field modulation, such a model presents a varied number of and positions for the quantum phase transition points. The critical exponents of the several quantum phase transition points were studied in more detail in Ref. \cite{Derzhko2}. There, the authors pointed out that although most of the critical points remain in the square lattice Ising model universality class - with exponents $\beta=0.125$, $\nu=1$, $\eta=0.25$, $\alpha=0$ and $\nu z=1$, one of them is characterized by a distinct set of exponents - namely $\beta=0.25$, $\nu=2$, $\eta=0.25$, $\alpha=-2$ and $\nu z=2$. In addition, they present the behavior of the inverse correlation length in the vicinity of some critical points. 

The effects of this regular alternation in the transverse field were also studied in the $XY$ model. A detailed discussion about its diagonalization, eigenvalue spectrum, and thermodynamic properties was given in \cite{Perk}. In \cite{Dutta}, a quenching in the anisotropy parameter along a critical surface was considered in a study of the dynamics of the one-dimensional $XY$ model in the presence of this period-2 transverse field. In \cite{Chanda1} this model was considered in the analysis of static and dynamical characteristics of nearest-neighbor entanglement, while in \cite{Chanda2} the same authors studied the model's entanglement emergence with temperature increase. 

\begin{figure}[h!]
 \centering
  \includegraphics[valign=c,scale=0.7]{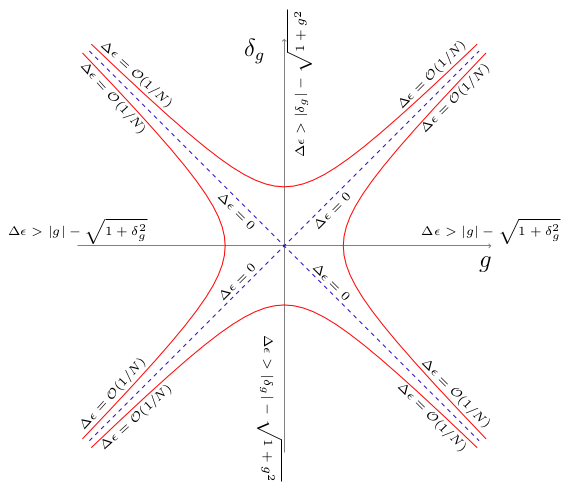}
  \includegraphics[valign=c,scale=0.5]{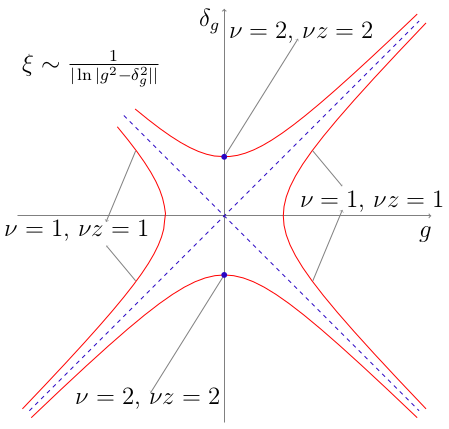} 
\caption{\label{Diagram}Diagrams summarizing our results. The transition phase lines are represented by the red hyperbolas in both charts. Shown on the left is the gap behavior in the various regions of the parameter $|g^2-\delta_g^2|$. Along these hyperbolas, the gap closes as $\Delta\epsilon=\mathcal{O}(1/N)$ for large $N$. External to them are the paramagnetic phase regions, where the gap never closes. Internal to the hyperbolas we have the Ising phase region, where the gap closes as $\Delta\epsilon=\mathcal{O}(\exp(-N/\xi)/\sqrt{N})$, when $N\gg\xi$, except along the blue dashed lines, where $g^2=\delta_g^2$ and the ground state is exactly degenerate. On the right we give the general expression for the correlation of the model and present the $\nu$ and $z$ critical exponents calculated in \cite{Derzhko2}. Along the hyperbolas these coefficients are given by $\nu=\nu z=1$, except on the blue dots, where $\delta_g^2=1$ and the system is characterized by a distinct set of exponents, with $\nu=\nu z=2$.}
\end{figure}

In this work we consider a spin-1/2 quantum Ising model in a transverse field with period-2 alternation and a finite number of spins $N$. Its Hamiltonian may be written as
\begin{equation}\label{Hamiltonian}
\hat H=-\frac{1}{2}\sum_{j=1}^N\bigg(\sigma_j^x\sigma_{j+1}^x+\big(g-(-1)^j\delta_g\big)\sigma_j^z\bigg),
\end{equation}
where $\sigma_j^a$, for $a=x,y,z$, are the Pauli spin operators of site $j$, subjected to periodic boundary conditions: $\sigma_{N+1}^a=\sigma_1^a$. We take $N/4\in\mathbb{N}$ for simplicity. As stated in Ref. \cite{Derzhko}, this model has critical points at $|g^2-\delta_g^2|=1$ and exhibits a paramagnetic phase for $|g^2-\delta_g^2|>1$ and a so-called Ising phase for $|g^2-\delta_g^2|<1$. The results found in the present work are depicted in Fig. \ref{Diagram}. 

The gap behavior may be divided into four regions. For $g^2=\delta_g^2$, the transverse field is zero for all even or all odd sites and the gap vanishes $\Delta\epsilon=0$. Along the phase transition lines, where $|g^2-\delta_g^2|=1$, the gap is bounded by functions of the inverse of the system size and closes as $\Delta\epsilon=\mathcal{O}(1/N)$ for large $N$. For $|g^2-\delta_g^2|>1$, corresponding to the paramagnetic region, the gap satisfies $\Delta\epsilon>\sqrt{G^2}-\sqrt{1+f^2}$, where $G^2=\max\{g^2,\delta_g^2\}$ and $f^2=\min\{g^2,\delta_g^2\}$, and never closes. Finally, in the Ising phase region, for $|g^2-\delta_g^2|<1$, we have that $\Delta\epsilon\sim|g^2-\delta_g^2|^{\frac{N}{2}}$ and the gap vanishes exponentially as $N$ increases. We also suggest a general expression for the correlation length of the model $\xi\sim1/|\ln|g^2-\delta_g^2||$, which is consistent with the numerical results given in \cite{Derzhko2} for $\delta_g=1$ and $g\rightarrow0,\,\sqrt{2}$.

In what follows we diagonalize the Hamiltonian \eqref{Hamiltonian}, determine its ground state, and obtain an expression for the energy gap between the different parity subspaces. Then we analyze its behavior in the various regimes of the field $g$ and system size $N$. In analogy with Ref. \cite{IsingModel}, we also suggest a general expression for the correlation length of the model that is consistent with the results obtained in Ref. \cite{Derzhko2} for its behavior in the vicinity of the critical points considered there.

\section{Diagonalizing the Hamiltonian}

In order to diagonalize the Hamiltonian \eqref{Hamiltonian}, we initially perform the Jordan-Wigner transformation \cite{Japas,qpt,Lieb}
\begin{equation}
 \begin{aligned}
  &\sigma_{2j+1}^+=\hat a_{2j+1}^{\dagger}\exp\bigg[\imath\pi\sum_{l=1}^j\hat b_{2l}^{\dagger}\hat b_{2l}+\imath\pi\sum_{l=0}^{j-1}\hat a_{2l+1}^{\dagger}\hat a_{2l+1}\bigg],\\
  \\[-1em]
  &\sigma_{2j}^+=\hat b_{2j}^{\dagger}\exp\bigg[\imath\pi\sum_{l=1}^{j-1}\hat b_{2l}^{\dagger}\hat b_{2l}+\imath\pi\sum_{l=1}^{j}\hat a_{2l-1}^{\dagger}\hat a_{2l-1}\bigg],
 \end{aligned}
\end{equation}
where $\sigma^{\pm}=(\sigma^x\pm\imath\sigma^y)/2$, $\hat a_j$ and $\hat b_ j$ are fermionic operators satisfying the canonical anticommutation relations $\{\hat a_i,\hat a_j^{\dagger}\}=\{\hat b_i,\hat b_j^{\dagger}\}=\delta_{ij}$, and all others anticommutators vanish. This distinction among fermionic operators for even and odd sites is made to address  the periodic alternation in the transverse field.

Hence, we get
\begin{equation}
 \begin{aligned}
  &\sigma_{2j}^z=\hat b_{2j}^{\dagger}\hat b_{2j}-\hat b_{2j}\hat b_{2j}^{\dagger},&&
 \sigma_{2j-1}^z=\hat a_{2j-1}^{\dagger}\hat a_{2j-1}-\hat a_{2j-1}\hat a_{2j-1}^{\dagger},\\
 \\[-1em]
   &\sigma_{2j-1}^x\sigma_{2j}^x=(\hat a_{2j-1}^{\dagger}-\hat a_{2j-1})(\hat b_{2j}^{\dagger}+\hat b_{2j}), &&
   \sigma_{2j}^x\sigma_{2j+1}^x=(\hat b_{2j}^{\dagger}-\hat b_{2j})(\hat a_{2j+1}^{\dagger}+\hat a_{2j+1}),\\
   \\[-1em]
   &\sigma_N^x\sigma_{N+1}^x=-(\hat b_N^{\dagger}-\hat b_N)(\hat a_1^{\dagger}+\hat a_1)\hat P,
 \end{aligned} 
\end{equation}
where the parity operator \cite{IsingModel}
\begin{equation}
 \hat P=\prod_{l=1}^{N/2}\bigg[\big(\mathds{1}-2\hat a_{2l-1}^{\dagger}\hat a_{2l-1}\big)\big(\mathds{1}-2\hat b_{2l}^{\dagger}\hat b_{2l}\big)\bigg]=\prod_{l=1}^{N/2}\big(\sigma_{2l-1}^z\sigma_{2l}^z\big),
\end{equation}
has eigenvalue $+1$ for states with an even number of down spins (or total number of fermions) and eigenvalue $-1$ for states with an odd number of down spins (or total number of fermions). From these definitions, we note that a state with a fermion occupying the mode $j$ corresponds to a state with a $z$-component up spin at site $j$. 

Thence, after some work, the Hamiltonian may be rewritten as
\begin{equation*}
 \begin{aligned}
  \hat H&=-\frac{1}{2}\bigg\{\sum_{j=1}^{N/2-1}\big[(\hat a_{2j-1}^{\dagger}-\hat a_{2j-1})(\hat b_{2j}^{\dagger}+\hat b_{2j})+(\hat b_{2j}^{\dagger}-\hat b_{2j})(\hat a_{2j+1}^{\dagger}+\hat a_{2j+1})\big]\\
  &\quad\quad\quad\quad\quad\quad+(\hat a_{N-1}^{\dagger}-\hat a_{N-1})(\hat b_N^{\dagger}+\hat b_N)-(\hat b_N^{\dagger}-\hat b_N)(\hat a_1^{\dagger}+\hat a_1)\hat P\\
  &\quad\quad\quad+\sum_{j=1}^{N/2}\big[\big(g+\delta_g\big)\big(\hat a_{2j-1}^{\dagger}\hat a_{2j-1}-\hat a_{2j-1}\hat a_{2j-1}^{\dagger}\big)+\big(g-\delta_g\big)\big(\hat b_{2j}^{\dagger}\hat b_{2j}-\hat b_{2j}\hat b_{2j}^{\dagger}\big)\big]\bigg\}.
 \end{aligned}
\end{equation*}

Let us define $\hat P^{\pm}=\frac{1}{2}\big(\mathds{1}\pm\hat P\big)$. The operators $\hat P^+$ and $\hat P^-$ are projectors onto the subspaces of positive (eigenvalue $+1$) and negative (eigenvalue $-1$) parity, respectively. Since $\hat P$ and $\hat H$ commute, they have simultaneous eigenstates and the latter may be diagonalized independently in each parity subspace. Therefore, we get $\hat H=\hat H^{+}\hat P^{+}+\hat H^{-}\hat P^{-}$, where
\begin{equation}
 \begin{aligned}
  \hat H^{\pm}&=-\frac{1}{2}\sum_{j=1}^{N/2}\bigg\{\big[(\hat a_{2j-1}^{\dagger}-\hat a_{2j-1})(\hat b_{2j}^{\dagger}+\hat b_{2j})+(\hat b_{2j}^{\dagger}-\hat b_{2j})(\hat a_{2j+1}^{\dagger}+\hat a_{2j+1})\big]\\
  \\[-1em]
  &\quad+\big(g+\delta_g\big)\big(\hat a_{2j-1}^{\dagger}\hat a_{2j-1}-\hat a_{2j-1}\hat a_{2j-1}^{\dagger}\big)+\big(g-\delta_g\big)\big(\hat b_{2j}^{\dagger}\hat b_{2j}-\hat b_{2j}\hat b_{2j}^{\dagger}\big)\bigg\},
 \end{aligned}
\end{equation}
with $\hat a_{N+1}=-\hat a_1$ for $\hat H^+$ and $\hat a_{N+1}=\hat a_1$ for $\hat H^-$.

\subsection{Positive parity subspace and even $N/2$:}
For the positive parity subspace and even $N/2$, we introduce the Fourier transformation
\begin{equation}\label{Fouriertrans}
\begin{aligned}
 &\hat a_{2j-1}=\frac{1}{\sqrt{N}}\sum_{k}(\hat a_k+\hat b_{k})e^{\imath k(2j-1)},\quad\hat b_{2j}=\frac{1}{\sqrt{N}}\sum_{k}(\hat a_k-\hat b_{k})e^{\imath k(2j)},\\
 \\[-1em]
 &k=\pm\frac{\pi}{N},\pm\frac{3\pi}{N},\pm\frac{5\pi}{N},\ldots,\pm\Big(\frac{\pi}{2}-\frac{\pi}{N}\Big),
\end{aligned}
\end{equation}
such that $\hat a_{N+1}=-\hat a_1$, with $\hat a_k$ and $\hat b_k$ fermionic operators. From this transformation, we observe that
\begin{equation}
 \sum_{j=1}^{N/2}\big(\hat a_{2j-1}^{\dagger}\hat a_{2j-1}+\hat b_{2j}^{\dagger}\hat b_{2j}\big)=\sum_k\big(\hat a_{k}^{\dagger}\hat a_{k}+\hat b_{k}^{\dagger}\hat b_{k}\big),
\end{equation}
which means that this transformation preserves the total number of particles and parity.

After some manipulations, we may write the positive-parity Hamiltonian as
\begin{equation}
 \hat H^+(\text{even } N/2)=\sum_{0<k<\pi/2}\hat A_k^{\dagger}\hat H_k\hat A_k,
\end{equation}
where
\begin{equation}\label{matrices}
%\begin{aligned}
   \hat A_k=\left[ \begin{array}{c}
               \hat a_k^{\dagger} \\
               \hat a_{-k} \\
               \hat b_{-k} \\
               \hat b_{k}^{\dagger}
              \end{array}\right],
 \text{ }
  \hat H_k=\left[ \begin{array}{cccc}
          g+\cos k & -\imath\sin k & 0 & \delta_g \\
          \imath\sin k & -(g+\cos k) & -\delta_g & 0\\
          0 & -\delta_g & -(g-\cos k) & -\imath\sin k\\
          \delta_g & 0 & \imath\sin k & (g-\cos k)
         \end{array}\right].
%\end{aligned}
\end{equation}

Thereof, we can obtain the spectrum of $\hat H^+(\text{even } N/2)$ by diagonalizing the matrix $\hat H_k$. It has eigenvalues $\pm\epsilon_k^{\pm}$, and $\hat H^+(\text{even } N/2)$ may be expressed in the form \cite{Perk,Dutta,Deng}
\begin{equation}\label{eigen}
\begin{aligned}
  &\hat H^+(\text{even } N/2)=\sum_{-\pi/2<k<\pi/2}\sum_{\nu=\pm}\epsilon_k^{\nu}\Big(\hat\gamma_{k,\nu}^{\dagger}\hat\gamma_{k,\nu}-1/2\Big),\\
  \\[-1em]
  &\epsilon_k^{\pm}=\sqrt{1+g^2+\delta_g^2\pm2\sqrt{g^2\delta_g^2+g^2\cos^2k+\delta_g^2\sin^2k}},
\end{aligned}
\end{equation}
where $\hat\gamma_{k,\nu}^{\dagger}$ is the fermionic quasiparticle creation operator associated with the mode $(k,\nu)$, with $\nu=+,-$. Accordingly, the ground state of $\hat H^+(\text{even } N/2)$ is given by the vacuum state of $\{\hat\gamma_{k,\nu}\}$. It can be checked that this state may be expressed in the form $\arrowvert\Psi\rangle_{\text{ground}}=\prod_{k>0}\arrowvert0\rangle_k$, with
\begin{equation}\label{groundstate}
 \arrowvert0\rangle_k=\big(\alpha_1^k+\alpha_2^k\,\hat b_{k}^{\dagger}\hat b_{-k}^{\dagger}+\alpha_3^k\,\hat a_{k}^{\dagger}\hat a_{-k}^{\dagger}+\alpha_4^k\,\hat a_{k}^{\dagger}\hat b_{-k}^{\dagger}+\alpha_5^k\,\hat a_{-k}^{\dagger}\hat b_{k}^{\dagger}+\alpha_6^k\,\hat a_{k}^{\dagger}\hat a_{-k}^{\dagger}\hat b_{k}^{\dagger}\hat b_{-k}^{\dagger}\Big)\arrowvert\text{vac}\rangle,
\end{equation}
where $\arrowvert\text{vac}\rangle$ is the state annihilated by all $\hat a_k$ and $\hat b_k$ operators. The complex coefficients $\alpha_i^k$ are determined in the diagonalization of $\hat H_k$ and $\sum_{i}|\alpha_i^k|=1$. The state \eqref{groundstate} contains an even number of particles and belongs to the positive-parity subspace. This ground-state energy is given by
\begin{equation}\label{PPGSEq}
 \epsilon^+(\text{even } N/2)=-\frac{1}{2}\sum_{-\pi/2<k<\pi/2}\big(\epsilon_k^++\epsilon_k^-\big).
\end{equation}

\subsection{Negative parity subspace and even $N/2$:}
For the negative parity subspace (with even $N/2$), we introduce the Fourier transformation
\begin{equation}
\begin{aligned}
 &\hat a_{2j-1}=\frac{1}{\sqrt{N}}\sum_{k}(\hat a_k+\hat b_{k})e^{\imath k(2j-1)},\quad\hat b_{2j}=\frac{1}{\sqrt{N}}\sum_{k}(\hat a_k-\hat b_{k})e^{\imath k(2j)},\\
 \\[-1em]
 &k=0,\pm\frac{2\pi}{N},\pm\frac{4\pi}{N},\ldots,\pm\Big(\frac{\pi}{2}-\frac{2\pi}{N}\Big),\frac{\pi}{2},
\end{aligned}
\end{equation}
such that $\hat a_{N+1}=\hat a_1$, with $\hat a_k$ and $\hat b_k$ fermionic operators. Again,
\begin{equation}
 \sum_{j=1}^{N/2}\big(\hat a_{2j-1}^{\dagger}\hat a_{2j-1}+\hat b_{2j}^{\dagger}\hat b_{2j}\big)=\sum_k\big(\hat a_{k}^{\dagger}\hat a_{k}+\hat b_{k}^{\dagger}\hat b_{k}\big),
\end{equation}
and the total number of particles and parity are conserved by this transformation.

Similarly to the previous case, we may write the negative-parity Hamiltonian as
\begin{equation}
 \hat H^-(\text{even } N/2)=\sum_{0<k<\pi/2}\hat A_k^{\dagger}\hat H_k\hat A_k+\frac{1}{2}\hat A_0^{\dagger}\hat H_0\hat A_0+\frac{1}{2}\hat A_{\frac{\pi}{2}}^{\dagger}\hat H_{\frac{\pi}{2}}\hat A_{\frac{\pi}{2}},
\end{equation}
where $\hat A_k$ and $\hat H_k$ are given by Eq. \eqref{matrices} and
\begin{equation}\label{matrices2}
\begin{aligned}
 &  \hat A_0^{\dagger}=\left[ \begin{array}{cccc}
               \hat a_0 & \hat a_0^{\dagger} & \hat b_0^{\dagger} & \hat b_0\\
              \end{array}\right],
 \quad
   \hat A_{\frac{\pi}{2}}^{\dagger}=\left[ \begin{array}{cccc}
               \hat a_{\frac{\pi}{2}} & \hat a_{\frac{\pi}{2}}^{\dagger} & \hat b_{\frac{\pi}{2}}^{\dagger} & \hat b_{\frac{\pi}{2}}\\
              \end{array}\right],\\
 \\[-1em]
 & \hat H_0=\left[ \begin{array}{cccc}
          g+1 & 0 & 0 & \delta_g \\
          0 & -(g+1) & -\delta_g & 0\\
          0 & -\delta_g & -(g-1) & 0\\
          \delta_g & 0 & 0 & g-1
         \end{array}\right],
 \quad
  \hat H_{\frac{\pi}{2}}=\left[ \begin{array}{cccc}
          g & 0 & \imath & \delta_g \\
          0 & -g & -\delta_g & \imath\\
          -\imath & -\delta_g & -g & 0\\
          \delta_g & -\imath & 0 & g
         \end{array}\right].
\end{aligned}
\end{equation}

Therefrom, we can obtain the spectrum of $\hat H^-(\text{even } N/2)$ by diagonalizing the matrices $\hat H_k$, $\hat H_0$, and $\hat H_{\frac{\pi}{2}}$. As we have seen before, $\hat H_k$ has eigenvalues $\pm \epsilon_k^{\pm}$ [see Eq. \eqref{eigen}] and the $k\setminus\{0,\pi/2\}$ part of the negative-parity Hamiltonian becomes
\begin{equation}
 \sum_{0<k<\frac{\pi}{2}}\hat A_k^{\dagger}\hat H_k\hat A_k=\sum_{\overset{-\pi/2<k<\pi/2}{k\neq0}}\sum_{\nu=\pm}\epsilon_k^{\nu}\Big(\hat\gamma_{k,\nu}^{\dagger}\hat\gamma_{k,\nu}-1/2\Big).
\end{equation}

The eigenvalues of $\hat H_0$ are $\pm\epsilon_0^{\pm}$, where $\epsilon_0^{\pm}=g\pm\sqrt{1+\delta_g^2}$. Defining $g_0=\sqrt{1+\delta_g^2}$, we have
\begin{equation}
  \hat A_0^{\dagger}\hat H_0\hat A_0=(g+g_0)(\hat\gamma_0^{\dagger}\hat\gamma_0-\hat\gamma_0\hat\gamma_0^{\dagger})+(g-g_0)(\hat\eta_0^{\dagger}\hat\eta_0-\hat\eta_0\hat\eta_0^{\dagger}),
\end{equation}
where
\begin{equation*}
 \begin{aligned}
  &\hat\gamma_0=\frac{1}{\sqrt{2g_0}}\big(\sqrt{g_0+1}\,\hat a_0+\sqrt{g_0-1}\,\hat b_0\big),\\
  &\hat\eta_0=\frac{1}{\sqrt{2g_0}}\big(\sqrt{g_0-1}\,\hat a_0-\sqrt{g_0+1}\,\hat b_0\big),
 \end{aligned}
\end{equation*}
are fermionic operators.

For $\hat H_{\frac{\pi}{2}}$, the eigenvalues are $\pm\epsilon_{\frac{\pi}{2}}^{\pm}$, where $\epsilon_\frac{\pi}{2}^{\pm}=\sqrt{1+g^2}\pm\delta_g$. Defining $g_{\frac{\pi}{2}}=\sqrt{1+g^2}$, we have
\begin{equation}
 \hat A_{\frac{\pi}{2}}^{\dagger}\hat H_{\frac{\pi}{2}}\hat A_{\frac{\pi}{2}}=(g_{\frac{\pi}{2}}+\delta_g)(\hat\gamma_{\frac{\pi}{2}}^{\dagger}\hat\gamma_{\frac{\pi}{2}}-\hat\gamma_{\frac{\pi}{2}}\hat\gamma_{\frac{\pi}{2}}^{\dagger})+(g_{\frac{\pi}{2}}-\delta_g)(\hat\eta_{\frac{\pi}{2}}^{\dagger}\hat\eta_{\frac{\pi}{2}}-\hat\eta_{\frac{\pi}{2}}\hat\eta_{\frac{\pi}{2}}^{\dagger}),
\end{equation}
where
\begin{equation*}
 \begin{aligned}
  &\hat\gamma_{\frac{\pi}{2}}=\frac{1}{2\sqrt{g_{\frac{\pi}{2}}}}\Big[\sqrt{g_{\frac{\pi}{2}}-g}\big(-\hat a_{\frac{\pi}{2}}+\hat b_{\frac{\pi}{2}}\big)+\imath\sqrt{g_{\frac{\pi}{2}}+g}\big(\hat a_{\frac{\pi}{2}}^{\dagger}+\hat b_{\frac{\pi}{2}}^{\dagger}\big)\Big],\\
  &\hat\eta_{\frac{\pi}{2}}=\frac{1}{2\sqrt{g_{\frac{\pi}{2}}}}\Big[\imath\sqrt{g_{\frac{\pi}{2}}-g}\big(\hat a_{\frac{\pi}{2}}+\hat b_{\frac{\pi}{2}}\big)+\sqrt{g_{\frac{\pi}{2}}+g}\big(-\hat a_{\frac{\pi}{2}}^{\dagger}+\hat b_{\frac{\pi}{2}}^{\dagger}\big)\Big],
 \end{aligned}
\end{equation*}
are again fermionic operators.

Therefore, $\hat H^-(\text{even } N/2)$ may be written in the form
\begin{equation}
\begin{aligned}
  \hat H^-(\text{even } N/2)=&\sum_{\overset{-\frac{\pi}{2}<k<\frac{\pi}{2}}{k\neq0}}\sum_{\nu=\pm}\epsilon_k^{\nu}\Big(\hat\gamma_{k,\nu}^{\dagger}\hat\gamma_{k,\nu}-1/2\Big)\\
  &+\frac{1}{2}\big[(g+g_0)(\hat\gamma_0^{\dagger}\hat\gamma_0-\hat\gamma_0\hat\gamma_0^{\dagger})+(g-g_0)(\hat\eta_0^{\dagger}\hat\eta_0-\hat\eta_0\hat\eta_0^{\dagger})\big]\\
  &+\frac{1}{2}\Big[(g_{\frac{\pi}{2}}+\delta_g)(\hat\gamma_{\frac{\pi}{2}}^{\dagger}\hat\gamma_{\frac{\pi}{2}}-\hat\gamma_{\frac{\pi}{2}}\hat\gamma_{\frac{\pi}{2}}^{\dagger})+(g_{\frac{\pi}{2}}-\delta_g)(\hat\eta_{\frac{\pi}{2}}^{\dagger}\hat\eta_{\frac{\pi}{2}}-\hat\eta_{\frac{\pi}{2}}\hat\eta_{\frac{\pi}{2}}^{\dagger})\Big].
\end{aligned}
\end{equation}

In order to determine the ground state of $\hat H^-(\text{even } N/2)$ we need to remember that it must have a negative parity. This means we ought to have an odd number of excited modes in it \cite{IsingModel}. The lowest energy is obtained when we have an $\eta$ particle in mode $0$ and all other modes are empty (see Appendix \ref{appendix1}). The ground-state energy of $\hat H^-(\text{even } N/2)$ is hence given by
\begin{equation}\label{NPGSEq}
 \epsilon^-(\text{even } N/2)=-\frac{1}{2}\sum_{\overset{-\pi/2<k<\pi/2}{k\neq0}}(\epsilon_k^++\epsilon_k^-)-\sqrt{1+g^2}-\sqrt{1+\delta_g^2}.
\end{equation}

\subsection{Positive parity subspace and odd $N/2$}
In this case we use the same Fourier transformation \eqref{Fouriertrans} but with momenta
\begin{equation}
 k=\pm\frac{\pi}{N},\pm\frac{3\pi}{N},\pm\frac{5\pi}{N},\ldots,\pm\Big(\frac{\pi}{2}-\frac{2\pi}{N}\Big), \frac{\pi}{2}.
\end{equation}

The positive parity Hamiltonian then becomes
\begin{equation}
 \hat H^+(\text{odd }N/2)=\sum_{0<k<\pi/2}\hat A_k^{\dagger}\hat H_k\hat A_k+\frac{1}{2}\hat A_{\frac{\pi}{2}}^{\dagger}\hat H_{\frac{\pi}{2}}\hat A_{\frac{\pi}{2}},
\end{equation}
where $\hat A_k$, $\hat H_k$ and $\hat A_{\frac{\pi}{2}}$, $\hat H_{\frac{\pi}{2}}$ are given by Eqs. \eqref{matrices} and \eqref{matrices2}, respectively. In diagonal form,
\begin{equation}
\begin{aligned}
  \hat H^+(\text{odd }N/2)=&\sum_{-\frac{\pi}{2}<k<\frac{\pi}{2}}\sum_{\nu=\pm}\epsilon_k^{\nu}\Big(\hat\gamma_{k,\nu}^{\dagger}\hat\gamma_{k,\nu}-1/2\Big)\\
  &+\frac{1}{2}\Big[(g_{\frac{\pi}{2}}+\delta_g)(\hat\gamma_{\frac{\pi}{2}}^{\dagger}\hat\gamma_{\frac{\pi}{2}}-\hat\gamma_{\frac{\pi}{2}}\hat\gamma_{\frac{\pi}{2}}^{\dagger})+(g_{\frac{\pi}{2}}-\delta_g)(\hat\eta_{\frac{\pi}{2}}^{\dagger}\hat\eta_{\frac{\pi}{2}}-\hat\eta_{\frac{\pi}{2}}\hat\eta_{\frac{\pi}{2}}^{\dagger})\Big].
\end{aligned}
\end{equation}

The ground-state is achieved when all modes $\gamma_{k,\nu;\,\frac{\pi}{2}}$ and $\eta_{\frac{\pi}{2}}$ are empty and the ground state energy in this case is
\begin{equation}\label{PPGSEqodd}
 \epsilon^+(\text{odd }N/2)=-\frac{1}{2}\sum_{-\pi/2<k<\pi/2}(\epsilon_k^++\epsilon_k^-)-\sqrt{1+g^2}.
\end{equation}

\subsection{Negative parity subspace and odd $N/2$}
In this case the momenta appearing in the Fourier transformation are
\begin{equation}
 k=0,\pm\frac{2\pi}{N},\pm\frac{4\pi}{N},\ldots,\pm\Big(\frac{\pi}{2}-\frac{\pi}{N}\Big).
\end{equation}

The negative parity Hamiltonian then becomes
\begin{equation}
 \hat H^-(\text{odd }N/2)=\sum_{0<k<\pi/2}\hat A_k^{\dagger}\hat H_k\hat A_k+\frac{1}{2}\hat A_{0}^{\dagger}\hat H_{0}\hat A_{0},
\end{equation}
where again $\hat A_k$, $\hat H_k$ and $\hat A_{\frac{\pi}{2}}$, $\hat H_{\frac{\pi}{2}}$ are those in Eqs. \eqref{matrices} and \eqref{matrices2}, respectively. Written in diagonal form,
\begin{equation}
\begin{aligned}
  \hat H^-(\text{odd }N/2)=&\sum_{-\frac{\pi}{2}<k<\frac{\pi}{2}}\sum_{\nu=\pm}\epsilon_k^{\nu}\Big(\hat\gamma_{k,\nu}^{\dagger}\hat\gamma_{k,\nu}-1/2\Big)\\
  &+\frac{1}{2}\big[(g+g_0)(\hat\gamma_0^{\dagger}\hat\gamma_0-\hat\gamma_0\hat\gamma_0^{\dagger})+(g-g_0)(\hat\eta_0^{\dagger}\hat\eta_0-\hat\eta_0\hat\eta_0^{\dagger})\big].
\end{aligned}
\end{equation}

The ground state is achieved when all modes $\gamma_{k,\nu;\,0}$ are empty and the mode $\eta_{0}$ is occupied. The ground-state energy in this case is
\begin{equation}\label{NPGSEqodd}
 \epsilon^-(\text{odd }N/2)=-\frac{1}{2}\sum_{\overset{-\pi/2<k<\pi/2}{k\neq0}}(\epsilon_k^++\epsilon_k^-)-\sqrt{1+\delta_g^2}.
\end{equation}

We are now in position to determine the ground-state energy of the whole Hamiltonian by comparing the ground-state energies of the positive- and negative-parity subspaces in the next section.

\section{Ground State and Gap Analysis}

Let us expand $\epsilon_k^{\pm}$ in the Fourier series $\epsilon_k^+=\sum_{l=0}^{\infty}u_l\cos(2kl)$ and $\epsilon_k^-=\sum_{l=0}^{\infty}v_l\cos(2kl)$. After some manipulations (see Appendix \ref{appendix2}) we get
\begin{equation}\label{gapcoef}
 \begin{aligned}
  \Delta\epsilon&=\text{sign}\big[(g^2-\delta_g)^{\frac{N}{2}}\big]\Theta\big(|\delta_g|-\sqrt{1+g^2}\big)\big(|\delta_g|-\sqrt{1+g^2}\big)\\
  \\[-1.5em]
  &+\Theta\big(|g|-\sqrt{1+\delta_g^2}\big)\big(|g|-\sqrt{1+\delta_g^2}\big)-\frac{N}{2}\sum_{n=0}^{\infty}\Big(u_{(2n+1)N/2}+v_{(2n+1)N/2}\Big),
 \end{aligned}
\end{equation}
where $\Delta\epsilon=\epsilon^--\epsilon^+$, $\Theta(x)$ is the Heaviside step function, and $\text{sign}(x)=x/|x|$. This equation holds for any $N$ even and for any $g$ and $\delta_g$.

In order to determine the sum in Eq. \eqref{gapcoef}, we note that $\epsilon_k^++\epsilon_k^-=\sqrt{(\epsilon_k^++\epsilon_k^-)^2}$, and, therefore,
\begin{equation*}\label{coef}
\begin{aligned}
  u_l+v_l&=\frac{2}{\pi}\int_{-\frac{\pi}{2}}^{\frac{\pi}{2}}\mathrm{d}k\cos(2kl)(\epsilon_k^++\epsilon_k^-)\\
  &=\frac{\sqrt{2}}{\pi}\int_{-\pi}^{\pi}\mathrm{d}k\cos(kl)\bigg[1+g^2+\delta_g^2+\sqrt{\sin^2k+\big[(g^2-\delta_g^2)-\cos k\big]^2}\bigg]^{\frac{1}{2}},
\end{aligned}
\end{equation*}
where we changed variables from $2k$ to $k$ in the last equality.

The cases $g^2-\delta_g^2=0$, $|g^2-\delta_g^2|=1$, and $0\neq|g^2-\delta_g^2|\neq1$ must be treated separately. For the latter we introduce $z=e^{\imath k}$ to obtain
\begin{equation}
 u_l+v_l=-\frac{\imath\sqrt{2}}{\pi}\oint_{|z|=1}\mathrm{d}zz^{l-1}\Bigg\{1+g^2+\delta_g^2+\sqrt{\frac{[z-(g^2-\delta_g^2)][1-(g^2-\delta_g^2)z]}{z}}\Bigg\}^{\frac{1}{2}}.
\end{equation}

\begin{figure}[h!]
\centering
  \includegraphics[width=0.7\textwidth]{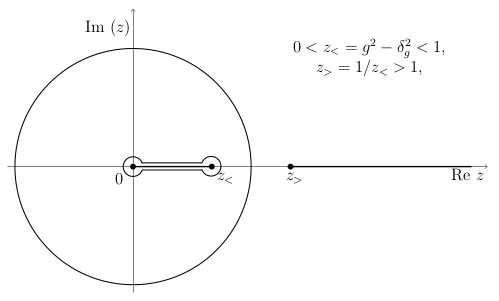}
\caption{\label{branchcuts}Illustration of the branch cuts and integration path used to obtain Eq. \eqref{eq1}.}
\end{figure}

To solve this integral, we first consider $0<g^2-\delta_g^2<1$. Then we make branch cuts along $(0,g^2-\delta_g^2)\cup(1/(g^2-\delta_g^2),\infty)$ (see Fig. \ref{branchcuts}) and by deforming the integration contour, we get
\begin{equation*}
 u_l+v_l=\frac{\imath2\sqrt{2}}{\pi}\int_0^1\mathrm{d}t(g^2-\delta_g^2)^lt^{2l-1}\Big(\sqrt{Z(t)}-\sqrt{\overbar{Z(t)}}\Big),
\end{equation*}
where $Z(t)=1+g^2+\delta_g^2+\imath\sqrt{(1-t^2)[1-(g^2-\delta_g^2)^2t^2]/t^2}$, and $\overbar{Z(t)}$ is its complex conjugate. Since Re$(Z)>0$ and Im$(Z)>0$, $\sqrt{Z(t)}-\sqrt{\overbar{Z(t)}}=\imath\sqrt{2}\sqrt{|Z|-\mathrm{Re}(Z)}$. Therefore,
\begin{equation}\label{eq1}
\begin{aligned}
  u_l+v_l&=-\frac{4}{\pi}(g^2-\delta_g^2)^l\times\\
  &\int_0^1\mathrm{d}t\,t^{2l-1}\Bigg[\sqrt{\frac{(1-t^2)[1-(g^2-\delta_g^2)^2t^2]}{t^2}+(1+g^2+\delta_g^2)^2}-(1+g^2+\delta_g^2)\Bigg]^{\frac{1}{2}}.
\end{aligned}
\end{equation}

For the other ranges of parameters, we follow the same procedure and derive very similar results. For $-1<g^2-\delta_g^2<0$ we get almost identical coefficients, with only the roles of $g^2$ and $\delta_g^2$ interchanged: $g^2\leftrightarrow\delta_g^2$. In the case that $g^2-\delta_g^2>1$, the coefficients will be the same except for the substitution of $(g^2-\delta_g^2)^l$ by $1/(g^2-\delta_g^2)^l$ outside the integral and of $[1-(g^2-\delta_g^2)^2t^2]$ by $[(g^2-\delta_g^2)^2-t^2]$ inside the square root in the integrand of Eq. \eqref{eq1}. In contrast, for $g^2-\delta_g^2<-1$, the coefficients differ from the ones for $g^2-\delta_g^2>1$ again just by the switch between $g^2$ and $\delta_g^2$.

Next we consider the particular cases $g^2-\delta_g^2=0$ and $g^2-\delta_g^2=\pm1$. In the case of the latter, we only take limits of $g^2-\delta_g^2\rightarrow\pm1$ in Eq. \eqref{eq1} and in the set of equations obtained from it according to the instructions in the preceding paragraph. For the former case we have $u_l+v_l=(2/\pi)\int_{-\pi}^{\pi}\mathrm{d}k\cos(kl)\sqrt{1+(g^2+\delta_g^2)/2}=4\sqrt{1+(g^2+\delta_g^2)/2}\delta_{l,0}$ and therefore the sum in Eq. \eqref{gapcoef} vanishes.

Using these results for the coefficients $u_l+v_l$ in the sum in Eq. \eqref{gapcoef} and that $\sum_{n=0}^{\infty}x^n=1/(1-x)$, we may summarize the gap between positive- and negative-parity ground states as
\begin{equation}\label{gapIsing}
 \begin{aligned}
 \Delta\epsilon&=\frac{1}{2}(g^2-\delta_g^2)^\frac{N}{2}\int_0^1\frac{\mathrm{d}t\quad t^{N-1}}{1-(\sqrt{|g^2-\delta_g^2|}t)^{2N}}\frac{4N}{\pi}\times\qquad\qquad\qquad\qquad\text{ }\qquad\qquad\\
 &\times\Bigg[\sqrt{\frac{(1-t^2)[1-|g^2-\delta_g^2|^2t^2]}{t^2}+(1+g^2+\delta_g^2)^2}-(1+g^2+\delta_g^2)\Bigg]^{\frac{1}{2}}%\\
 %\\[-1em]
 %&
 \,\text{for }|g^2-\delta_g^2|<1,
 \\
 \end{aligned}
\end{equation} 
\begin{equation}\label{gapPara}
 \begin{aligned}
\Delta\epsilon&=\text{sign}\big[(g^2-\delta_g)^{\frac{N}{2}}\big]\Theta\big(|\delta_g|-\sqrt{1+g^2}\big)\big(|\delta_g|-\sqrt{1+g^2}\big)\\
  \\[-1.5em]
  &+\Theta\big(|g|-\sqrt{1+\delta_g^2}\big)\big(|g|-\sqrt{1+\delta_g^2}\big)+\frac{1}{2(g^2-\delta_g^2)^\frac{N}{2}}\int_0^1\frac{\mathrm{d}t\quad t^{N-1}}{1-(t/\sqrt{|g^2-\delta_g^2|})^{2N}}\frac{4N}{\pi}\\
 &\times\Bigg[\sqrt{\frac{(1-t^2)[|g^2-\delta_g^2|^2-t^2]}{t^2}+(1+g^2+\delta_g^2)^2}-(1+g^2+\delta_g^2)\Bigg]^{\frac{1}{2}}%\\
 %\\[-1em]
 %&
 \text{ for }|g^2-\delta_g^2|>1,
 \end{aligned}
\end{equation}
\begin{equation}\label{gapTransition}
 \quad\Delta\epsilon=\text{sign}\big[(g^2-\delta_g^2)^\frac{N}{2}\big]\frac{1}{2}\int_0^1\mathrm{d}t\,\frac{4N}{\pi}\frac{t^{N-3/2}}{1-t^{2N}}\sqrt{\sqrt{(1-t^2)^2+(2G^2t)^2}-2G^2t},\qquad\qquad\qquad\quad
\end{equation}
where $G^2=g^2$ for $g^2-\delta_g^2=1$, $G^2=\delta_g^2$ for $g^2-\delta_g^2=-1$, and                                                                                                                            
\begin{equation}\label{NoGap}
\begin{aligned}
 &\Delta\epsilon=0\qquad,&&\text{ for } g^2-\delta_g^2=0.
\end{aligned}
\end{equation}

The exact degeneracy in Eq. \eqref{NoGap} is explained by a the symmetry acquired by the Hamiltonian \eqref{Hamiltonian} when $g^2=\delta_g^2$. When $\delta_g=g$, the transverse field vanishes in even sites and we have $[\hat H,\,\sigma_{2j}^{x}]=0$. However, this operator anticommutes with the parity operator: $\hat P\sigma_{2j}^{x}=-\sigma_{2j}^{x}\hat P$. Therefore, for a state with definite parity $p$ and eigenenergy $E_0$, $\arrowvert p,\,E_0\rangle$, we have $\hat H(\sigma_{2j}^{x}\arrowvert p,\,E_0\rangle)=E_0(\sigma_{2j}^{x}\arrowvert p,\,E_0\rangle)$ and $\hat P(\sigma_{2j}^x\arrowvert p,\,E_0\rangle)=-p(\sigma_{2j}^{x}\arrowvert p,\,E_0\rangle)$. Then, if $\arrowvert p,\,E_0\rangle$ is observed to be a ground state of the Hamiltonian, the state with opposite parity, $\sigma_{2j}^{x}\arrowvert p,\,E_0\rangle$, is also a possible ground state. In the case that $\delta_g=-g$, the transverse field vanishes in odd sites and $[\hat H,\,\sigma_{2j+1}^{x}]=0$ while $\hat P\sigma_{2j+1}^x=-\sigma_{2j+1}^x\hat P$, which leads to this same degeneracy.

Expressions \eqref{gapIsing}-\eqref{gapTransition} allow us to conclude that the Hamiltonian \eqref{Hamiltonian} always has a positive-parity ground state, except for when $g^2-\delta_g^2<0$ and $N/2$ is odd. Thus, for $N/2$ even, the ground-state energy is given by equation Eq. \eqref{PPGSEq}. When $N/2$ is odd, the ground-state energy is given by Eq. \eqref{PPGSEqodd}, if $g^2>\delta_g^2$, or by Eq. \eqref{NPGSEqodd}, if $g^2<\delta_g^2$.

We notice that in the paramagnetic phase \eqref{gapPara}, the gap is always manifest, meaning that it does not close in the thermodynamic limit $N\rightarrow\infty$. On the other hand, Eqs. \eqref{gapTransition} and \eqref{gapIsing} are bounded by functions of $N$, as we show in the sequence.

In the case of Eq. \eqref{gapTransition}, taking from now on the parity of the ground state to always be positive (for the negative parity ground state all relations below hold with the switch $\Delta\epsilon\rightarrow-\Delta\epsilon$), we have (see Appendix \ref{appendix3})
\begin{equation}\label{BoundGapTransition}
\begin{aligned}
 \frac{1}{2\sqrt{G^2}}\bigg[\tan\Big(\frac{\pi}{4N}\Big)+\frac{\pi}{12N}\bigg]
 \leqslant\Delta\epsilon\leqslant\tan\Big(\frac{\pi}{4N}\Big).
\end{aligned}
\end{equation}
This shows that, when $N$ is large,
\begin{equation}
 \Delta\epsilon=\mathcal{O}(1/N),\quad \text{for }|g^2-\delta_g^2|=1.
\end{equation}

In the case of the so-called Ising phase \eqref{gapIsing}, we have again (see Appendix \ref{appendix3})
\begin{equation}\label{BoundGapIsing}
\begin{aligned}
&\frac{1}{2\sqrt{2(1+g^2+\delta_g^2)}}\max\Bigg\{|g^2-\delta_g^2|^{\frac{N}{2}}\,\frac{2}{\sqrt{\pi}}\frac{\sqrt{1-|g^2-\delta_g^2|}}{\sqrt{N}},|g^2-\delta_g^2|^{\frac{N}{2}}\,\frac{4\sqrt{|g^2-\delta_g^2|}}{\pi N}\Bigg\}\\
\\
&\qquad\qquad\qquad\qquad\qquad\qquad\qquad\qquad\leqslant\Delta\epsilon\leqslant\\
\\
&\frac{\sqrt{2(1+|g^2-\delta_g^2|)}}{2}\Bigg[|g^2-\delta_g^2|^{\frac{N}{2}}\,\frac{\pi\sqrt{|g^2-\delta_g^2|}}{2N-1}+|g^2-\delta_g^2|^{\frac{N}{2}}\,2\frac{\sqrt{1-|g^2-\delta_g^2|}}{\sqrt{N-1}}\Bigg].
\end{aligned}
\end{equation}
These results allow us to conclude that the gap in the Ising phase is exponentially small in $N$ and vanishes in the thermodynamic limit, as one would expect from the results in Ref. \cite{Derzhko}. Furthermore, in Ref. \cite{Hatano} the authors stated that there is a relation between the gap and correlation length for this type of system of the kind $\Delta\epsilon\sim\exp(-N/\xi)$. This result and the similarity of the content of Eq. \eqref{BoundGapIsing} to that in Eq. ($18$), as well as the discussion following it in Ref. \cite{IsingModel}, suggest that the correlation length of the model is given by
\begin{equation}\label{correlength}
 \xi\sim1/|\ln|g^2-\delta_g^2||.
\end{equation}
We state that this relation is \emph{suggested} because, to our knowledge, it is not mathematically established under what conditions the aforementioned relation between the gap and correlation length is valid.

The behavior of the correlation length of this model in the vicinity of several critical points was investigated in Ref. \cite{Derzhko2}. There, the authors found that for $\delta_g=1$, quantum phase transition points occur at $g^*=0$ and $g^*=\sqrt{2}$ and that the correlation length dependence on $g$ around these points was given by $\xi\sim1/g^2$ and $\xi\sim1/|\sqrt{2}-g|$, respectively. Using $\delta_g=1$ in Eq. \eqref{correlength}, we get $\xi\sim1/|\ln|g^2-1||$. Expanding this function around $0$, for $g\rightarrow0$, and around $\sqrt{2}$, for $g\rightarrow\sqrt{2}$, we get precisely the results stated above for the dependence of the correlation length of the model on the applied field. This also advocates for the suggestion of this relation. 

From this we may conclude that, as in the homogeneous case, when $\xi\ll N$, the gap closes as $\Delta\epsilon=\mathcal{O}(\exp(-N/\xi)/\sqrt{N})$. While, when $\xi>N$, the gap closes as $\Delta\epsilon=\mathcal{O}(1/N)$, in the same way as in the critical behavior.

\section{Conclusion}

In this work we have studied the finite-size spin-1/2 quantum Ising model with a period-2 alternating transverse field. Similarly to its homogeneous counterpart, this model exhibits quantum phase transitions, with the distinction that the number and positions of the quantum phase transition points are depended on the transverse field modulation. We showed that the Hamiltonian for such a model may be broken into two orthogonal subspaces containing an even or odd number of quasiparticles introduced by a Jordan-Wigner transformation or, equivalently, containing an even or odd number of down spins. We determined the ground-state energy for each subspace separately and then derived closed-form expressions for the energy gap between them. Therefrom, we obtained the ground-state energy for the whole Hamiltonian and analyzed the dependence of the gap on the system size in the different quantum phases and in the quantum phase transition. We see that, in general, the results are analogous to those found for the model's homogeneous counterpart. Furthermore, we suggested a general expression for the correlation length of this model in terms of the transverse field. The relation suggested is consistent with previous results found for the behavior of the correlation length in the vicinity of the distinct critical points characterized by different critical exponents that emerge in this system. We stress that, although in principle this quantity could be analytically calculated this has not yet been done.

\section*{Acknowledgements}
We acknowledge financial support from the Brazilian agencies Conselho Nacional de Desenvolvimento Cient\'ifico e Tecnol\'ogico (CNPq) and Coordenação de Aperfeiçoamento de Pessoal de N\'ivel Superior (CAPES).

% ELEMENTOS PÓS-TEXTUAIS
% ----------------------------------------------------------
% Apêndices
% ----------------------------------------------------------
\appendix

\section{Negative Parity Ground State}\label{appendix1}

Here we show that the ground-state energy of the negative-parity subspace is truly the one in Eq. \eqref{NPGSEq}. We ought to have an odd number of excited modes in such a ground state \cite{IsingModel}. If the excited mode is in the $k\setminus\{0,\pi/2\}$ part of $\hat H^-(\text{even } N/2)$, supposing it is in the mode $(k^{'},-)$ that gives the lowest value of $\epsilon_k^-$, the global state may be written as 
\begin{equation}
\arrowvert\Psi_1^-\rangle=\prod_{k>k^{'}}\arrowvert0\rangle_k\otimes\arrowvert01\rangle_{k^{'}}\otimes\prod_{0<k<k^{'}}\arrowvert0\rangle_k\otimes\arrowvert0\rangle_{0,\,\frac{\pi}{2}}                                                                                                                                                                                                                                                                                                                                                                                                                                                                                                                                                                                                                                                                                                                                                                                                           \end{equation}
where $\arrowvert0\rangle_k$ is given by Eq. \eqref{groundstate}, $\arrowvert01\rangle_{k^{'}}=\gamma_{k^{'},-}^{\dagger}\arrowvert0\rangle_k$\footnote{The $0$ in $\arrowvert01\rangle_{k^{'}}$ is to indicate that the modes $(\pm k^{'},+)$ and $(-k^{'},-)$ are empty.} 
and $\arrowvert0\rangle_{0,\frac{\pi}{2}}=\arrowvert0\rangle_0\otimes\arrowvert0\rangle_{\frac{\pi}{2}}$ is the state annihilated by all operators $\hat x_n$, for $x=\gamma,\,\eta$ and $n=0,\,\frac{\pi}{2}$. Its energy will be given by
\begin{equation}
 E_1=-\frac{1}{2}\sum_{\overset{-\pi/2<k<\pi/2}{k\neq0,k^{'}}}(\epsilon_k^++\epsilon_k^-)-\frac{1}{2}\epsilon_{k^{'}}^++\frac{1}{2}\epsilon_{k^{'}}^--g-g_{\frac{\pi}{2}}.
\end{equation}
We note that this is the lowest energy possible for such a ground state, since $\epsilon_{k^{'}}^+>\epsilon_{k^{'}}^-$, and exciting other modes would only increase the energy.

Otherwise, if the excited mode(s) are in the $\{0,\,\pi/2\}$ part of $\hat H^-(\text{even } N/2)$, the global state will be given by 
\begin{equation}
 \arrowvert\Psi_2^-\rangle=\prod_{k>0}\arrowvert0\rangle_k\otimes\arrowvert\psi\rangle_{0,\frac{\pi}{2}},
\end{equation}
with $\arrowvert\psi\rangle_{0,\frac{\pi}{2}}$ one of the possibilities
\begin{equation}\label{possibles}
\begin{aligned}
 \big\{&\arrowvert1000\rangle,\,\arrowvert0100\rangle,\,\arrowvert1011\rangle,\,\arrowvert0111\rangle,\,\arrowvert0010\rangle,\,\arrowvert0001\rangle,\,\arrowvert1110\rangle,\,\arrowvert1101\rangle\big\},
\end{aligned}
\end{equation}
where
\begin{equation}
\begin{aligned}
&\arrowvert1000\rangle=\gamma_0^{\dagger}\arrowvert0\rangle_{0,\frac{\pi}{2}},\qquad\qquad\qquad\,\arrowvert0100\rangle=\eta_0^{\dagger}\arrowvert0\rangle_{0,\frac{\pi}{2}},\\
&\arrowvert0010\rangle=\gamma_{\frac{\pi}{2}}^{\dagger}\arrowvert0\rangle_{0,\frac{\pi}{2}},\qquad\qquad\qquad\arrowvert0001\rangle=\eta_{\frac{\pi}{2}}^{\dagger}\arrowvert0\rangle_{0,\frac{\pi}{2}}. 
\end{aligned}
\end{equation}
In terms of modes $a$ and $b$, we have
\begin{equation}
\begin{aligned}
  &\arrowvert0\rangle_0=\arrowvert0\rangle_0^a\arrowvert0\rangle_0^b,\\
  &\arrowvert1000\rangle=\frac{1}{\sqrt{2g_0}}\big(\sqrt{g_0+1}\,\arrowvert1\rangle_0^a\arrowvert0\rangle_0^b+\sqrt{g_0-1}\,\arrowvert0\rangle_0^a\arrowvert1\rangle_0^b\big)\otimes\arrowvert0\rangle_{\frac{\pi}{2}},\\
  &\arrowvert0100\rangle=\frac{1}{\sqrt{2g_0}}\big(\sqrt{g_0-1}\,\arrowvert1\rangle_0^a\arrowvert0\rangle_0^b-\sqrt{g_0+1}\,\arrowvert0\rangle_0^a\arrowvert1\rangle_0^b\big)\otimes\arrowvert0\rangle_{\frac{\pi}{2}},\\
  &\arrowvert1100\rangle=\arrowvert1\rangle_0^a\arrowvert1\rangle_0^b\otimes\arrowvert0\rangle_{\frac{\pi}{2}}
\end{aligned}
\end{equation}
and
\begin{equation}
 \begin{aligned}
  &\arrowvert0\rangle_{\frac{\pi}{2}}=\frac{1}{\sqrt{2g_{\frac{\pi}{2}}}}\big(\sqrt{g_{\frac{\pi}{2}}-g}+\imath\sqrt{g_{\frac{\pi}{2}}+g}\,\hat a_{\frac{\pi}{2}}^{\dagger}\hat b_{\frac{\pi}{2}}^{\dagger}\big)\arrowvert0\rangle_{\frac{\pi}{2}}^a\arrowvert0\rangle_{\frac{\pi}{2}}^b,\\
  &\arrowvert0010\rangle=\arrowvert0\rangle_0\otimes\frac{1}{\sqrt{2}}\big(\arrowvert0\rangle_{\frac{\pi}{2}}^a\arrowvert1\rangle_{\frac{\pi}{2}}^b-\arrowvert1\rangle_{\frac{\pi}{2}}^a\arrowvert0\rangle_{\frac{\pi}{2}}^b\big),\\
  &\arrowvert0001\rangle=\arrowvert0\rangle_0\otimes\frac{-\imath}{\sqrt{2}}\big(\arrowvert0\rangle_{\frac{\pi}{2}}^a\arrowvert1\rangle_{\frac{\pi}{2}}^b+\arrowvert1\rangle_{\frac{\pi}{2}}^a\arrowvert0\rangle_{\frac{\pi}{2}}^b\big),\\
  &\arrowvert0011\rangle=\arrowvert0\rangle_0\otimes\frac{1}{\sqrt{2g_{\frac{\pi}{2}}}}\big(\sqrt{g_{\frac{\pi}{2}}+g}-\imath\sqrt{g_{\frac{\pi}{2}}-g}\,\hat a_{\frac{\pi}{2}}^{\dagger}\hat b_{\frac{\pi}{2}}^{\dagger}\big)\arrowvert0\rangle_{\frac{\pi}{2}}^a\arrowvert0\rangle_{\frac{\pi}{2}}^b,
\end{aligned}
\end{equation}
where $\hat x_n\arrowvert0\rangle_n^x=0$ and $\arrowvert1\rangle_n^x=\hat x_n^{\dagger}\arrowvert0\rangle_n^x$, with $n=0,\,\frac{\pi}{2}$ and $x=a,\,b$. This confirms that the states in Eq. \eqref{possibles} belongs to the negative parity subspace.

We set $\hat H_{0,\frac{\pi}{2}}$ equal to
\begin{equation}
\hat H_{0,\frac{\pi}{2}}=\frac{1}{2}\Big[\epsilon_{\frac{\pi}{2}}^{+}(\hat\gamma_{\frac{\pi}{2}}^{\dagger}\hat\gamma_{\frac{\pi}{2}}-\hat\gamma_{\frac{\pi}{2}}\hat\gamma_{\frac{\pi}{2}}^{\dagger})+\epsilon_{\frac{\pi}{2}}^{-}(\hat\eta_{\frac{\pi}{2}}^{\dagger}\hat\eta_{\frac{\pi}{2}}-\hat\eta_{\frac{\pi}{2}}\hat\eta_{\frac{\pi}{2}}^{\dagger})+\epsilon_0^+(\hat\gamma_0^{\dagger}\hat\gamma_0-\hat\gamma_0\hat\gamma_0^{\dagger})+\epsilon_0^-(\hat\eta_0^{\dagger}\hat\eta_0-\hat\eta_0\hat\eta_0^{\dagger})\Big],
\end{equation}
where we recall that $\epsilon_0^{\pm}=g\pm g_0$ and $\epsilon_{\frac{\pi}{2}}^{\pm}=g_{\frac{\pi}{2}}\pm\delta_g$. Thus, we have
\begin{equation}
 \begin{aligned}
  &\hat H_{0,\frac{\pi}{2}}\arrowvert0010\rangle=(\delta_g-g)\arrowvert0010\rangle,%\\&
  &&\qquad\hat H_{0,\frac{\pi}{2}}\arrowvert0001\rangle=-(g+\delta_g)\arrowvert0001\rangle,\\
  &\hat H_{0,\frac{\pi}{2}}\arrowvert1110\rangle=(g+\delta_g)\arrowvert1110\rangle,%\\&
  &&\qquad\hat H_{0,\frac{\pi}{2}}\arrowvert1101\rangle=-(\delta_g-g)\arrowvert1101\rangle,\\
  &\hat H_{0,\frac{\pi}{2}}\arrowvert1000\rangle=(g_0-g_{\frac{\pi}{2}})\arrowvert1000\rangle,%\\&
  &&\qquad\hat H_{0,\frac{\pi}{2}}\arrowvert0100\rangle=-(g_0+g_{\frac{\pi}{2}})\arrowvert0100\rangle,\\
  &\hat H_{0,\frac{\pi}{2}}\arrowvert1011\rangle=(g_0+g_{\frac{\pi}{2}})\arrowvert1011\rangle,%\\
  &&\qquad\hat H_{0,\frac{\pi}{2}}\arrowvert0111\rangle=(-g_0+g_{\frac{\pi}{2}})\arrowvert0111\rangle.
 \end{aligned}
\end{equation}
Among all these possibilities, the one with the lowest energy is the state $\arrowvert0100\rangle$ - i.e. the state in which there is an $\eta$ particle in mode $0$ while all others are empty. Therefore, the ground-state energy would be
\begin{equation}
 E_2=-\frac{1}{2}\sum_{\overset{-\pi/2<k<\pi/2}{k\neq0}}(\epsilon_k^++\epsilon_k^-)-g_0-g_{\frac{\pi}{2}}.
\end{equation}

Now, using that $g_0=\sqrt{1+\delta_g^2}$,
\begin{equation}
 E_2-E_1=-g_0-g_{\frac{\pi}{2}}-\epsilon_{k^{'}}^-+g+g_{\frac{\pi}{2}}=-(\sqrt{1+\delta_g^2}-g)-\epsilon_{k^{'}}^-.
\end{equation}
We have that
\begin{equation}
 \frac{d}{dk}\epsilon_k^-=\frac{1}{2\epsilon_k^-}\frac{(g^2-\delta_g^2)\sin(2k)}{\sqrt{g^2\delta_g^2+g^2\cos^2k+\delta_g^2\sin^2k}}\begin{cases}
               >0,& \text{ if } k>0\\
               <0,& \text{ if } k<0                                                                                                                   
             \end{cases}\text{ and } g^2>\delta_g^2,
\end{equation}
which means that $\epsilon_{k}^-$ has a global minimum at $k=0$, for $g^2>\delta_g^2$, and then\footnote{Note that we explicitly have $\epsilon_0^{\pm}\neq\epsilon_{k=0}^{\pm}$ and $\epsilon_{\frac{\pi}{2}}^{\pm}\neq\epsilon_{k=\frac{\pi}{2}}^{\pm}$.}
\begin{equation}
\begin{aligned}
 &g-\sqrt{1+\delta_g^2}-\epsilon_{k^{'}}^-=\big||g|-\sqrt{1+\delta_g^2}\big|-\epsilon_{k^{'}}^-=\epsilon_{k=0}^--\epsilon_{k^{'}}^-<0,\text{ for }g\geqslant\sqrt{1+\delta_g^2}\wedge\forall k^{'},\\
 \\
 &-(\sqrt{1+\delta_g^2}-g)-\epsilon_{k^{'}}^-<0,\quad\text{ for }g\leqslant\sqrt{1+\delta_g^2}\wedge\forall k^{'}.
\end{aligned}
\end{equation}

Hence, $E_2<E_1$, $\forall g, \delta_g$ and the ground-state energy of $\hat H^-(\text{even } N/2)$ is given by Eq. \eqref{NPGSEq}.

\section{Gap}\label{appendix2}

The Fourier coefficients in the series expansions of $\epsilon_k^{\pm}$ are given by
\begin{equation}
 u_l=\frac{2}{\pi}\int_{-\frac{\pi}{2}}^{\frac{\pi}{2}}\mathrm{d}k\cos(2kl)\epsilon_k^+,\qquad
 v_l=\frac{2}{\pi}\int_{-\frac{\pi}{2}}^{\frac{\pi}{2}}\mathrm{d}k\cos(2kl)\epsilon_k^-.
\end{equation}
This way, using also that $\epsilon_{-k}^{\pm}=\epsilon_k^{\pm}$,
\begin{equation}
\begin{aligned}
 \epsilon^+(\text{even } N/2)&=-\sum_{l=0}^{\infty}(u_l+v_l)\sum_{\overset{0<k<\pi/2}{k=(2n+1)\pi/N}}\cos(2kl),\\
 \\[-1.5em]
 \epsilon^+(\text{odd } N/2)&=-\sum_{l=0}^{\infty}(u_l+v_l)\sum_{\overset{0<k<\pi/2}{k=(2n+1)\pi/N}}\cos(2kl)-\sqrt{1+g^2},\\
 \\[-1.5em]
 \epsilon^-(\text{even } N/2)&=-\sum_{l=0}^{\infty}(u_l+v_l)\sum_{\overset{0<k<\pi/2}{k=2n\pi/N}}\cos(2kl)-\sqrt{1+g^2}-\sqrt{1-\delta_g^2},\\
 \\[-1.5em]
 \epsilon^-(\text{odd } N/2)&=-\sum_{l=0}^{\infty}(u_l+v_l)\sum_{\overset{0<k<\pi/2}{k=2n\pi/N}}\cos(2kl)-\sqrt{1+\delta_g^2}.
\end{aligned}
\end{equation}

In the positive-parity subspace,
\begin{equation}
 \sum_{0<k<\pi/2}\cos(2kl)=\begin{cases}
                            \frac{N}{4}(-1)^n\delta_{l,nN/2}\qquad\qquad\quad\,\text{for }N/2\text{ even}\\
                            \frac{N}{4}(-1)^n\delta_{l,nN/2}-\frac{1}{2}(-1)^l\quad\text{for }N/2\text{ odd},                         \end{cases}\,n\in\mathbb{N}.
\end{equation}
where we made use of the relation $\sum_{s=0}^{m-1}\cos(x+sy)=\cos[x+(m-1)y/2]\sin(my/2)\csc(y/2)$ and
\begin{equation}
 \lim_{x\rightarrow nN/2}\frac{1}{2}\frac{\sin(\pi x)}{\sin(2\pi x/N)}=\delta_{l,nN/2}\frac{N}{4}\times\begin{cases}
                                                                                   (-1)^n\,\text{for }N/2\text{ even}\\
                                                                                   1\,\qquad\text{for }N/2\text{ odd},                                                                                  \end{cases}\text{with }n\in\mathbb{N}.
\end{equation}
Thus,
\begin{equation}\label{eigeven}
\begin{aligned}
 \epsilon^+(\text{even } N/2)=&-\frac{N}{4}\Big[\big(u_{0}+v_{0}\big)-\big(u_{N/2}+v_{N/2}\big)+\big(u_{N}+v_{N}\big)-\big(u_{3N/2}+v_{3N/2}\big)+\ldots\Big];\\
 \\[-1em]
 \epsilon^+(\text{odd }N/2)=&-\frac{N}{4}\Big[\big(u_{0}+v_{0}\big)-\big(u_{N/2}+v_{N/2}\big)+\big(u_{N}+v_{N}\big)-\big(u_{3N/2}+v_{3N/2}\big)+\ldots\Big]\\
 \\[-1.5em]
 &+\frac{1}{2}\Big[\big(u_0-u_1+u_2-u_3+\ldots\big)+\big(v_0-v_1+v_2-v_3+\ldots-2\sqrt{1+g^2}\big)\Big].
\end{aligned}
\end{equation}

Likewise, in the negative-parity subspace,
\begin{equation}
 \sum_{0<k<\pi/2}\cos(2kl)=\begin{cases}
                            \frac{N}{4}\delta_{l,nN/2}-\delta_{l,2n}\qquad\text{for }N/2\text{ even}\\
                            \frac{N}{4}\delta_{l,nN/2}-\frac{1}{2}\quad\qquad\text{for }N/2\text{ odd},                         \end{cases}\text{with }n\in\mathbb{N}.
\end{equation}
Thus,
\begin{equation}\label{eigodd}
\begin{aligned}
  \epsilon^-(\text{even } N/2)=&-\frac{N}{4}\Big[\big(u_{0}+v_{0}\big)+\big(u_{N/2}+v_{N/2}\big)+\big(u_{N}+v_{N}\big)+\big(u_{3N/2}+v_{3N/2}\big)+\ldots\Big]\\
  \\[-1.5em]
 &+\Big(u_0+u_2+u_4+\ldots-\sqrt{1+\delta_g^2}\Big)+\Big(v_0+v_2+v_4+\ldots-\sqrt{1+g^2}\Big),\\
 \\[-1em]
  \epsilon^-(\text{odd } N/2)=&-\frac{N}{4}\Big[\big(u_{0}+v_{0}\big)+\big(u_{N/2}+v_{N/2}\big)+\big(u_{N}+v_{N}\big)+\big(u_{3N/2}+v_{3N/2}\big)+\ldots\Big]\\
 \\[-1.5em]
 &+\frac{1}{2}\Big[\big(u_0+u_1+u_2+u_3+\ldots-2\sqrt{1+\delta_g^2}\big)+\big(v_0+v_1+v_2+v_3+\ldots\big)\Big].
\end{aligned}
\end{equation}
However,
\begin{equation}
 \begin{aligned}
  &\epsilon_{k=0}^+=\sum_{l=0}^{\infty}u_l\cos(0)=u_0+u_1+u_2+u_3+\ldots=\Big||g|+\sqrt{1+\delta_g^2}\Big|,\\
  &\epsilon_{k=\frac{\pi}{2}}^+=\sum_{l=0}^{\infty}u_l\cos(\pi l)=u_0-u_1+u_2-u_3+\ldots=\Big||\delta_g|+\sqrt{1+g^2}\Big|,\\
  &\epsilon_{k=0}^-=\sum_{l=0}^{\infty}v_l\cos(0)=v_0+v_1+v_2+v_3+\ldots=\Big||g|-\sqrt{1+\delta_g^2}\Big|,\\
  &\epsilon_{k=\frac{\pi}{2}}^-=\sum_{l=0}^{\infty}v_l\cos(\pi l)=v_0-v_1+v_2-v_3+\ldots=\Big||\delta_g|-\sqrt{1+g^2}\Big|.
 \end{aligned}
\end{equation}
Hence,
\begin{equation}\label{theta1}
 \begin{aligned}
  &\big(u_0+u_2+u_4+\ldots-\sqrt{1+\delta_g^2}\big)+\big(v_0+v_2+v_4+\ldots-\sqrt{1+g^2}\big)\\
  \\[-1.5em]
  &=\frac{1}{2}\Big(\Big||g|+\sqrt{1+\delta_g^2}\Big|+\Big||g|-\sqrt{1+\delta_g^2}\Big|\Big)-\sqrt{1+\delta_g^2}\\
  \\[-1.5em]
  &\,+\frac{1}{2}\Big(\Big||\delta_g|+\sqrt{1+g^2}\Big|+\Big||\delta_g|-\sqrt{1+g^2}\Big|\Big)-\sqrt{1+g^2}\\
  \\[-1.5em]
  &=\Theta\Big(|g|-\sqrt{1+\delta_g^2}\Big)\Big(|g|-\sqrt{1+\delta_g^2}\Big)+\Theta\Big(|\delta_g|-\sqrt{1+g^2}\Big)\Big(|\delta_g|-\sqrt{1+g^2}\Big)
 \end{aligned}
\end{equation}
and
\begin{equation}\label{theta2}
 \begin{aligned}
  &\big(u_1+u_3+u_5+\ldots-\sqrt{1+\delta_g^2}\big)+\big(v_1+v_3+v_5+\ldots+\sqrt{1+g^2}\big)=\\
  \\[-1.5em]
  &=\frac{1}{2}\Big(\Big||g|+\sqrt{1+\delta_g^2}\Big|+\Big||g|-\sqrt{1+\delta_g^2}\Big|\Big)-\sqrt{1+\delta_g^2}\\
  \\[-1.5em]
  &\,-\frac{1}{2}\Big(\Big||\delta_g|+\sqrt{1+g^2}\Big|+\Big||\delta_g|-\sqrt{1+g^2}\Big|\Big)+\sqrt{1+g^2}\\
  \\[-1.5em]
  &=\Theta\Big(|g|-\sqrt{1+\delta_g^2}\Big)\Big(|g|-\sqrt{1+\delta_g^2}\Big)-\Theta\Big(|\delta_g|-\sqrt{1+g^2}\Big)\Big(|\delta_g|-\sqrt{1+g^2}\Big),
 \end{aligned}
\end{equation}
where $\Theta(x)=\begin{cases}
                  1,& \text{ if } x>0\\
                  0,& \text{ if } x<0\text{ }
                 \end{cases}$ is the Heaviside step function.                
Thence, subtracting Eq. \eqref{eigeven} from Eq. \eqref{eigodd} and using Eqs. \eqref{theta1} and \eqref{theta2}, we obtain Eq. \eqref{gapcoef}.
    
\section{Bounds}\label{appendix3}

Here we show how one gets Eqs. \eqref{BoundGapTransition} and \eqref{BoundGapIsing}. We start with the former. We assume the parity of the ground state to be positive, but all relations derived here hold in the case of a ground state with negative parity with the change $\Delta\epsilon\rightarrow-\Delta\epsilon$. 

We first observe that Eq. \eqref{gapTransition} may be rewritten as
\begin{equation}
 \Delta\epsilon=\frac{1}{2}\int_0^1\mathrm{d}t\,\frac{4N}{\pi}\frac{t^{N-3/2}}{1-t^{2N}}\frac{1-t^2}{\sqrt{\sqrt{(1-t^2)^2+(2G^2t)^2}+2G^2t}}.
\end{equation}
The function $f(t)=\sqrt{\sqrt{(1-t^2)^2+(2G^2t)^2}+2G^2t}$ is monotonically increasing in the interval $t\in[0,1]$ for $G^2\geqslant1$ and hence
\begin{equation}
 \frac{1}{4\sqrt{G^2}}\int_0^1\mathrm{d}t\,\frac{4N}{\pi}\frac{t^{N-3/2}(1-t^2)}{1-t^{2N}}\leqslant\Delta\epsilon\leqslant\frac{1}{2}\int_0^1\mathrm{d}t\,\frac{4N}{\pi}\frac{t^{N-3/2}(1-t)}{1-t^{2N}},
\end{equation}
where we also used that $1-t^2\leqslant1-t$ to obtain the result on the right-hand side of the inequality. We have that
\begin{equation}
\begin{aligned}
\int_0^1\mathrm{d}t\,\frac{4N}{\pi}\frac{t^{N-3/2}(1-t)}{1-t^{2N}}&=2\tan\big(\frac{\pi}{4N}\big),\,\text{ and }\\
\\
\int_0^1\mathrm{d}t\,\frac{4N}{\pi}\frac{t^{N-1/2}(1-t)}{1-t^{2N}}&=\frac{2}{\pi}\Big[\psi^{(0)}\Big(\frac{1}{2}+\frac{3}{4N}\Big)-\psi^{(0)}\Big(\frac{1}{2}+\frac{1}{4N}\Big)\Big]\\
&=\frac{1}{\pi N}\sum_{n=0}^{\infty}\frac{1}{(n+\frac{1}{2}+\frac{3}{4N})(n+\frac{1}{2}+\frac{1}{4N})}\\
&>\frac{1}{\pi N}\sum_{n=0}^{\infty}\frac{1}{(n+1)^2}=\frac{\pi}{6N}, 
\end{aligned}
\end{equation}
where $\psi^{(0)}(z)=\frac{d}{dz}\ln\Gamma(z)$ is the digamma function \cite{abram}. This leads directly to Eq. \eqref{BoundGapTransition}.

To derive the inequality \eqref{BoundGapIsing} we closely follow the arguments in Appendix $A$ of Ref. \cite{IsingModel}. We have that
\begin{equation}
\begin{aligned}
  \Delta&\epsilon>-\frac{N}{2}\big(u_{N/2}+v_{N/2}\big)=\frac{1}{2}|g^2-\delta_g^2|^{\frac{N}{2}}\\
  &\int_0^1\mathrm{d}t\,\frac{4N}{\pi}t^{N-3/2}\big[\sqrt{(1-t^2)(1-|g^2-\delta_g^2|^2t^2)+(1+g^2+\delta_g^2)^2t^2}-(1+g^2+\delta_g^2)t\big]^{\frac{1}{2}}\\
  \\
  &=\int_0^1\frac{\mathrm{d}t}{\pi}\frac{t^{N-3/2}\,2N|g^2-\delta_g^2|^{\frac{N}{2}}\sqrt{(1-t^2)(1-|g^2-\delta_g^2|^2t^2)}}{\big[\sqrt{(1-t^2)(1-|g^2-\delta_g^2|^2t^2)+(1+g^2+\delta_g^2)^2t^2}+(1+g^2+\delta_g^2)t\big]^{\frac{1}{2}}}.
\end{aligned}
\end{equation}
The function in the denominator of the above integral is monotonically increasing in the interval $t\in[0,1]$ and hence
\begin{equation}
\begin{aligned}
  \Delta\epsilon&\geqslant\frac{1}{2\sqrt{2(1+g^2+\delta_g^2)}}|g^2-\delta_g^2|^{\frac{N}{2}}\int_0^1\mathrm{d}t\,\frac{4N}{\pi}t^{N-3/2}\sqrt{(1-t^2)(1-|g^2-\delta_g^2|^2t^2)}\\
  &\geqslant\frac{1}{2\sqrt{2(1+g^2+\delta_g^2)}}|g^2-\delta_g^2|^{\frac{N}{2}}\int_0^1\mathrm{d}t\,\frac{4N}{\pi}t^{N-3/2}\sqrt{(1-t)(1-|g^2-\delta_g^2|t)},
\end{aligned}
\end{equation}
where we used that $\sqrt{(1+t)(1+|g^2-\delta_g^2|t)}\geqslant1$ in the last inequality. This integral is identical to the one appearing in Ref \cite{IsingModel} in the calculation made in Appendix $A$ for the lower bound for the ferromagnetic gap, with the change of the parameter $g$ there by $|g^2-\delta_g^2|^{\frac{1}{2}}$ here. Thence, we just replicate their result here to get
\begin{equation}\label{lbound}
\begin{aligned}
  \Delta\epsilon\geqslant\frac{1}{2\sqrt{2(1+g^2+\delta_g^2)}}\max\Bigg\{&|g^2-\delta_g^2|^{\frac{N}{2}}\,\frac{2}{\sqrt{\pi}}\frac{\sqrt{1-|g^2-\delta_g^2|}}{\sqrt{N}},\\
 &|g^2-\delta_g^2|^{\frac{N}{2}}\,\frac{4\sqrt{|g^2-\delta_g^2|}}{\pi N}\Bigg\}.
\end{aligned}
\end{equation}

For the upper bound,
\begin{equation}
 \begin{aligned}
  &\Delta\epsilon=-\frac{N}{2}\sum_{n=0}^{\infty}\big(u_{(2n+1)N/2}+v_{(2n+1)N/2}\big)\\
  \\[-1em]
  &=\sum_{n=0}^{\infty}\int_0^1\frac{\mathrm{d}t}{\pi}\frac{t^{(2l+1)N-3/2}\,2N|g^2-\delta_g^2|^{\frac{(2l+1)N}{2}}\sqrt{(1-t^2)(1-|g^2-\delta_g^2|^2t^2)}}{\big[\sqrt{(1-t^2)(1-|g^2-\delta_g^2|^2t^2)+(1+g^2+\delta_g^2)^2t^2}+(1+g^2+\delta_g^2)t\big]^{\frac{1}{2}}}\\
  \\[-1em]
  &\leqslant\sqrt{\frac{1+|g^2-\delta_g^2|}{2}}\sum_{n=0}^{\infty}|g^2-\delta_g^2|^{\frac{(2l+1)N}{2}}\int_0^1\mathrm{d}t\,\frac{4N}{\pi}t^{(2l+1)N-3/2}\sqrt{(1-t)(1-|g^2-\delta_g^2|t)},
 \end{aligned}
\end{equation}
where we used the monotonicity of the function in the denominator of the integral in the second equality and that $\sqrt{(1+t)(1+|g^2-\delta_g^2|t)}\leqslant\sqrt{2(1+|g^2-\delta_g^2|)}$ to obtain the following inequality. Again, the sum appearing in this last line is identical to the one in Appendix $A$ of Ref. \cite{IsingModel} in the calculation of the upper bound of the ferromagnetic gap of the Ising model with $|g^2-\delta_g^2|^{\frac{1}{2}}$ replacing the role of the parameter $g$. Using the result found there,
\begin{equation}\label{ubound}
 \Delta\epsilon\leqslant \sqrt{\frac{1+|g^2-\delta_g^2|}{2}} \Bigg[|g^2-\delta_g^2|^{\frac{N}{2}}\,\frac{\pi\sqrt{|g^2-\delta_g^2|}}{2N-1}+|g^2-\delta_g^2|^{\frac{N}{2}}\,2\frac{\sqrt{1-|g^2-\delta_g^2|}}{\sqrt{N-1}}\Bigg].
\end{equation}

Combining Eqs. \eqref{lbound} and \eqref{ubound}, we obtain Eq. \eqref{BoundGapIsing}.
% ----------------------------------------------------------
% Referências bibliográficas
% ----------------------------------------------------------
\bibliographystyle{apsrev4-1}
\bibliography{references}

\end{document}